\documentstyle[12pt]{article}
\normalbaselines
\begin{document}

\baselineskip=24pt

\bibliographystyle{unsrt}
\vbox {\vspace{6mm}}

\begin{center} {\bf DIFFERENT REALIZATIONS OF 
TOMOGRAPHIC PRINCIPLE\\ 
IN QUANTUM STATE MEASUREMENT}
\end{center}

\bigskip

\begin{center} 

{\it Stefano Mancini }$^{\dag}$,  
{\it Vladimir I. Man'ko }$^{\ddag}$   
{\it and Paolo Tombesi }$^{\dag}$

\end{center}

\bigskip

\begin{center}

$^{\dag}$ Dipartimento di Matematica e Fisica, Universit\`a 
di Camerino, 
I-62032 Camerino, Italy\\
$^{\ddag}$ Lebedev Physical Institute, Leninsky Prospect 53, 
117924 
Moscow, Russia

\end{center}

\bigskip

\bigskip

\begin{abstract}

\baselineskip=24pt

We establish a general principle for the tomographic 
approach to quantum state 
reconstruction, till now  based on a simple rotation 
transformation in the 
phase space, which allows us to consider other types 
of transformations.
Then, we will present different realizations of the 
principle in 
specific examples.

\end{abstract}

\bigskip
\bigskip
\bigskip
\bigskip

\section{Introduction}\label{s1}

The tomography is well known in the field of medicine where 
it is extensively used for images 
reconstruction in diagnostic systems. 
It is based on the possibility of
recording transmission profiles of the radiation which has 
penetrated a 
living body from various directions. A collection of these 
distributions, by means of computer assisted mathematical 
procedures 
\cite{medicaltom}, 
allows to obtain the desired physical density distribution
giving insight into the body.

An approach, based on this principle, to get the Wigner 
function of a 
quantum system was first proposed by J. Bertrand and P. 
Bertrand \cite{ber}.

In quantum optics one has the opportunity of 
measuring all 
possible quadratures of an e.m. field by means of the homodyne 
detection,
so that the tomography can be easily implemented.
In fact, recently, Vogel and Risken \cite{vogel} pointed out that
the homodyne detected marginal distribution is just the Radon 
transform (or "tomography")  of the Wigner function.
By inverting the Radon tranform one can obtain the Wigner 
function and 
then recover the state; this is the basis of the method proposed by 
Smithey {\it et al.} \cite{raimer}. 

The density matrix elements, in some 
representations, can also be
obtained by avoiding the Wigner function and then the Radon 
transform 
\cite{wvogel,darianopar,APL}.
In particular in Ref. \cite{APL}, the system density operator was 
expressed as a convolution of the marginal distribution of the 
homodyne 
output and a kernel operator.

Anyway, also at quantum optical level the tomographic reconstruction 
relies on the possibility of performing measurements of observables 
obtained by 
means of transformations belonging to the group $O(2)$. 
Here, starting from this transformation, we would establish a general 
principle which include other types of transformations, 
i.e. other schemes 
for the quantum state measurement. This will be done in Sec. 2, while 
in the following we
shall consider different realizations of 
this principle. In particular, in Sec. 3 we present the symplectic 
tomography,
in Sec. 4 the latter is compared with the homodyne tomography 
and in Sec. 5 we discuss the photon number tomography.

\section{The tomographic principle}

In the usual Optical Homodyne Tomography the observed quantities are 
the quadratures ${\hat x}_{\phi}={\hat q}\cos\phi+{\hat p}\sin\phi$ 
obtained as mixtures of position ${\hat q}$ and momentum ${\hat p}$ 
by means of a rotation $g$ in the phase space
\begin{eqnarray}\label{grot}
\left (\begin{array}{c}
q\\
p\end{array}\right )\Rightarrow g
\left (\begin{array}{c}
q\\
p\end{array}\right );~~~~~~~
g=\left (\begin{array}{clcr}
\cos \,\phi &\sin \,\phi\\
-\sin \,\phi &\cos \,\phi \end{array}\right ).
\end{eqnarray}

The quadrature histograms $w(x,\phi)$, also called marginal 
distributions, 
are projections (Radon transformations) of the Wigner function 
\cite{Wig}
\begin{equation}\label{Wpro}
w(x,\phi)=\int W(q\cos\phi-p\sin\phi,
q\sin\phi+p\cos\phi)\,dp\,.
\end{equation}
From the set of histograms $w(x,\phi)$ the Wigner function itself 
(hence the quantum state) can be reconstructed via the inverse Radon 
transform as was shown in Ref. \cite{vogel}.

On the other hand, the marginal distribution $w(x,\phi)$ results as 
\cite{vogel}
\begin{equation}\label{wdef}
w(x,\phi)=\langle x_{\phi}|\hat\rho|x_{\phi}\rangle
=\langle q|{\cal G}(g)\hat\rho{\cal G}^{-1}(g)|q\rangle\,,
\end{equation}
where $|x_{\phi}\rangle$ are eigenkets of quadrature operators
and ${\cal G}(g)$ is the unitary group representation for the 
transformation $g$.
In this case
\begin{equation}\label{Grot}
{\cal G}(g)=\exp\left[i\phi\left(\frac{{\hat p}^2}{2}
+\frac{{\hat q}^2}{2}\right)\right]\,.
\end{equation}

Thus the above well known state reconstruction procedure could be 
generalized into the following principle:
given a density operator $\hat\rho$ 
and a group element $g$,
one can create different types of tomography if, 
by knowing the matrix 
elements 
$\langle x|{\cal G}(g)\hat\rho{\cal G}^{-1}(g)|x\rangle$,
from measurements, 
is able to invert the formula
expressing the density operator in terms of the above "marginal" 
distribution (the $x$ may denote either
continous or discrete eigenvalues). For the inversion procedure 
one can use the properties of
summation or integration over group parameters.
The only problem is mathematical 
one to make the inversion and/or
physical one to realize the transformation ${\cal G}(g)$ 
in laboratory.

It is worth noting that other tomographic-like approaches  
already known \cite{ulf,ole} can be taken back to this principle,
but we would now put our attention on two other particular 
realizations.

We would also remark that the marginal distribution function 
$w$ depends 
on one variable and is determined by some 
extra parameters.
Since we are treating the latter in the same way of variables, 
in the follows, we often use the steatment 
marginal distribution instead of
set of marginal distributions.

\section{Symplectic tomography}

\subsection{Formalism}

Let us consider the quadrature observable $\hat X$ as a 
generic linear form in position $\hat
q$ and momentum $\hat p$
\begin{equation}\label{quadef}
\hat X=\mu \hat q+\nu \hat p+\delta
\end{equation}
with $\mu,\nu,\delta$ real parameters (their physical meaning will 
be discussed later on), then
it is possible to get the density matrix elements from the marginal 
distribution, avoiding the
evaluation of the Wigner function as an intermediate step. For this 
pourpose we start from a well
known \cite{cahgla69} representation of the density operator  
\begin{equation}\label{rho1}
\hat\rho=\int \frac{dq\,dp}{2\pi} W(q,p)\hat T(q,p)
\end{equation}
where the Wigner function $W(q,p)$ is a weight function for the 
expansion of the density
operator in terms of the operator $\hat T(q,p)$, which is defined 
as the complex Fourier
transform of the displacement operator $\hat D(\xi)$ \cite{cahgla69}
\begin{equation}\label{Topdef}
{\hat T}(q,p)=\int\frac{dudv}{2\pi}
e^{i({\hat q}-q)v-i({\hat p}-p)u}\,.
\end{equation}

Following the lines of Ref. \cite{MMT}
we may write the marginal distribution $w$ for the generic
quadrature of Eq. (\ref{quadef}) as
\begin{equation}\label{margx}
w(x,\mu,\nu)=\int e^{-ik(x-\mu q-\nu p)}
W(q,p)\frac{dkdqdp}{(2\pi)^2}\,,
\end{equation}
with $x=X-\delta$.
By means of the Fourier transform of the function $w$ one 
can then obtain the relation
\begin{equation}\label{wig}
W(q,p)=(2\pi)^2z^2{\tilde w}(z,-zq,-zp)
\end{equation}
where $-zq,\,-zp,\,z$ are the conjugate variable to $\mu,\,\nu,\,x$
respectively and the Fourier transform $\tilde w$ has the property
\begin{equation}\label{wpro}
{\tilde w}(z,-zq,-zp)=\frac{1}{z^2}{\tilde w}(1,-q,-p)\,.
\end{equation}
It is worth remarking that in this case the connection between the 
Wigner function and the marginal
distribution is simply guaranteed by means of the Fourier transform 
instead of the Radon one. 

We now present a specific example of the measurable probability
distribution $w$; to this end we consider the Schr\"odinger cat 
state of the type discussed in Ref. \cite{vogel}, i.e.
\begin{equation}\label{psicat}
|\Psi\rangle=\frac{|a+ib\rangle+|a-ib\rangle}
{\{2[1+\cos(2ab)\exp(-2b^2)]\}^{1/2}}\,,
\end{equation}
with $a$ and $b$ arbitrary real numbers. Hence in this case we 
have \cite{reco}
\begin{eqnarray}\label{wcat}
w(x,\mu,\nu)&=&\left(\frac{2}{\pi}\right)^{1/2}
\left[\frac{1}{\mu^2+\nu^2}\right]^{1/2}
\frac{1}{[1+\cos(2ab)\exp(-2b^2)]}
\exp\left[-2\frac{(x-\mu a)^2+b^2\nu^2}{\mu^2+\nu^2}\right]
\nonumber\\
&\times&\left\{\cosh\left[\frac{4\nu b(x-\mu a)}{\mu^2+\nu^2}
\right]
+\cos\left[\frac{2b(2\mu x-a(\mu^2-\nu^2))}{\mu^2+\nu^2}\right]
\right\}\,.
\end{eqnarray}
The result of Ref. \cite{vogel} can be reproduced as partial case 
of this formula.

To get an invariant expression of the density operator 
in terms of 
the marginal distribution $w$, we insert Eq. (\ref{wig}) into Eq. 
(\ref{rho1}) and expressing the 
marginal distribution in terms of the
Fourier transform, we have:
\begin{equation}\label{denop}
\hat\rho=\int\frac{dqdp}{(2\pi)^2}\hat T(q,p)\int dx d\mu d\nu \quad
z^2w(x,\mu,\nu)e^{-izx+i\mu zq+i\nu zp},
\end{equation}
or, in a compact form 
\begin{equation}\label{density}
\hat\rho=\int dx d\mu d\nu \quad w(x,\mu,\nu)\hat K_{\mu,\nu}\,,
\end{equation}
where the kernel operator $\hat K_{\mu,\nu}$ is given by
\begin{equation}\label{kersym}
\hat K_{\mu,\nu}
=\frac{1}{2\pi}z^2e^{-izx}e^{-\frac{z}{\sqrt{2}}
(\nu-i\mu)\hat a^{\dag}}
e^{\frac{z}{\sqrt{2}}(\nu+i\mu)\hat a}
e^{-\frac{z^2}{4}(\mu^2+\nu^2)}\,.
\end{equation}
The fact that $\hat K_{\mu,\nu}$ depends on the $z$ variable as well 
(i.e. each Fourier component
gives a selfconsistent kernel) shows the overcompleteness of 
information 
achievable by measuring the
observable of Eq. (\ref{quadef}). 

If, and only if, the kernel operator is bounded every moment of the 
kernel is bounded for
all possible distributions $w(x,\mu,\nu)$, then, according to the 
central limit theorem, the
matrix elements of Eq. (\ref{density}) can be sampled on a 
sufficiently 
large set of data. This is only possible in
the number or coherent states basis, as can be evicted from the 
expressions (\ref{kersym}),
analogously to 
the results of
Ref. \cite{APL}. 

We named the developed procedure "Symplectic 
Tomography" \cite{reco},
since in this case the "marginal" distribution is obtained 
by using a 
symplectic transformation $g$ belonging to the symplectic group 
$ISp(2,R)$
\begin{eqnarray}\label{gsym}
\left (\begin{array}{c}
q\\
p\end{array}\right )\Rightarrow g
\left (\begin{array}{c}
q\\
p\end{array}\right );~~~~~~~
g=\left (\begin{array}{clcr}
\cos \,\phi &\sin \,\phi\\
-\sin \,\phi &\cos \,\phi \end{array}\right )
\left (\begin{array}{clcr}
\lambda &0\\
0&\lambda ^{-1}\end{array}\right ).
\end{eqnarray}
For this transformation one has
\begin{equation}\label{munusym}
\mu =\lambda \cos \,\phi;~~~\nu =\lambda ^{-1} \sin \,\phi;
~~~\delta =0.
\end{equation}
Thus, for the realization of the scheme, the element $g$  
is the product of squeezing and rotation operators. 
It means that for
our scheme the representation operator is
\begin{equation}\label{Gsym}
{\cal G}(g)=\exp \left [i\phi\left (\frac {\hat p^2}{2}+
\frac {\hat q^2}{2}\right )\right ]\exp \left [\frac {i\lambda }{2}
\left (\hat q\hat p+\hat p\hat q\right )\right ].
\end{equation}
A related tomographic approach based 
on symplectic group was formulated for short pulse 
measurements in Ref. \cite{josa}.
In the two dimensional case  
the symplectic group reduces to $SL(2,R)$, but
the formalism is still valid at higher dimensions for
transformations belonging to $Sp(2n,R)\;(n>1)$
\cite{reco},
and the scheme might be 
generalized to other Lie groups 
different from the symplectic one. 

\subsection{Applications}

The quadrature of Eq.
(\ref{quadef}) could be experimentally accessible by using 
for example the squeezed
pre-amplification (pre-attenuation) of a field mode which is 
going to be measured (a similar method
in different context was discussed in Ref. \cite{leopauPRL}). 
In fact, let
$\hat a$ be the signal field mode to be detected, when it 
passes through a squeezer it becomes
${\hat a}_s={\hat a}\cosh s-{\hat a}^{\dag}\sinh s$, 
where $s$ is
the squeezing parameter
\cite{LK}. Then, if we subsequently detect the field by using 
the balanced homodyne scheme, we get
an output signal proportional to the average of the following 
quadrature
\begin{equation}\label{N1-N2}
{\hat E}(\phi)=\frac{1}{\sqrt{2}}({\hat a}_se^{-i\phi}+{\hat
a}^{\dag}_se^{i\phi})\,,
\end{equation}
where $\phi$ is the local oscillator phase. Eq. 
(\ref{N1-N2}) can be rewritten as
\begin{equation}\label{quamea}
{\hat E}(\phi)=\frac{1}{\sqrt{2}}\left({\hat a}
[e^{-i\phi}\cosh s-e^{i\phi}\sinh s]
+{\hat a}^{\dag}[e^{i\phi}\cosh s-e^{-i\phi}\sinh s]\right)\,,
\end{equation}
which coincides with Eq. (\ref{quadef}) if one
recognizes the independent parameters
\begin{equation}\label{munumeas}
\mu=[\cosh s-\sinh s]\cos\phi;\quad\nu=
[\cosh s+\sinh s]\sin\phi\,.
\end{equation}
The shift parameter $\delta$ has not a real physical meaning, 
since it causes only a displacement of
the distribution along the $X$ line without changing its shape, 
as can be evicted from Eq.
(\ref{margx}). So, in a practical situation 
it can be omitted.
To be more precise, the shift parameter does not play a real 
physical role in the measurement process,
it has been introduced for formal completeness and it expresses 
the possibility to achieve the desired
marginal distribution by performing the measurements in an 
ensemble of frames which are  each
other shifted; (a related method was early discussed in Ref.
\cite{Royer1}). In an electro-optical system this only means 
to have the freedom of using different
photocurrent scales in which the zero is shifted by a known amount.
An experimental method based on an observable similar to that of 
Eq. (\ref{quadef}) was also proposed in Ref. \cite{Zucchetti}.

Our approach can be extended 
to multimode systems and,
in the two-mode case 
it is interesting to use the connection between the 
Wigner function
and the marginal distribution of only one quadrature for the 
case of heterodyne detection \cite{reco}
(particulary used to detect multimode squeezed states).

Beside that it is worth noting that the marginal distribution 
$w(x,\mu,\nu)$ introduced in the symplectic tomography consents an 
alternative formulation of the quantum dynamics in terms of 
classical 
probability distributions \cite{ourFofP}. In fact it has the 
characteristics of reality, positivity, normalizability and 
measurability.
Then, we can derive  the evolution equation 
for the marginal distribution function
$w$ using the invariant form of the connection between the 
marginal
distribution and the density operator, given by the formula 
(\ref{density}).
From the equation of
motion for the density operator 
\begin{equation}\label{rhoH}
\partial_t\hat\rho=-i[\hat H,\hat\rho]\,,
\end{equation}
we obtain the evolution equation for the marginal distribution 
in the form
\begin{equation}\label{wevo}
\int dxd\mu d\nu\; \left\{\dot w(x,\mu,\nu,t)\hat K_{\mu,\nu}+
w(x,\mu,\nu,t)\hat I_{\mu,\nu}\right\}=0
\end{equation}
in which the known Hamiltonian determines the kernel $\hat 
I_{\mu,\nu}$ 
through the
commutator
\begin{equation}\label{I}
\hat I_{\mu,\nu}=i[\hat H,\hat K_{\mu,\nu}]\,.
\end{equation}
The obtained integral-operator equation can be reduced to an 
integro-differential equation for
the function $w$ in some cases. To do this we 
represent the kernel operator
$\hat I_{\mu,\nu}$ in the normal order form (i.e. all the momentum 
operators on the left side and the
position ones on the right side) containing the operator 
$\hat K_{\mu,\nu}$ as follow
\begin{equation}\label{Inormal}
:\hat I_{\mu,\nu}:={\cal R}(\hat p):\hat K_{\mu,\nu}:{\cal P}
(\hat q)
\end{equation}
where ${\cal R}(\hat p)$ and ${\cal P}(\hat q)$ are finite or 
infinite operator polynomials
(depending also on the parameters $\mu$ and $\nu$) determined 
by the Hamiltonian.
Then, we calculate the matrix elements of the operator equation 
(\ref{wevo}) between the states 
$\langle p|$ and $|q\rangle$ obtaining
\begin{equation}\label{inteq}
\int dxd\mu d\nu\; \left\{\dot w(x,\mu,\nu,t)+
w(x,\mu,\nu,t){\cal R}(p){\cal P}(q)\right\}
\langle p|:\hat K_{\mu,\nu}:|q\rangle=0\,.
\end{equation}
If we suppose to write
\begin{equation}\label{RP}
{\cal R}(p){\cal P}(q)=\Pi(p,q)=\sum_n\sum_m c_{n,m}
(z,\mu,\nu)p^nq^m\,,
\end{equation}
due to the particular form of the kernel in Eq. (\ref{kersym}), 
Eq. (\ref{inteq}) can be rewritten
as
\begin{equation}\label{Piright}
\int dxd\mu d\nu\; \left\{\dot w(x,\mu,\nu,t)+
w(x,\mu,\nu,t){\overrightarrow\Pi}(\tilde p,\tilde q)\right\}
\langle p|:\hat K_{\mu,\nu}:|q\rangle=0\,,
\end{equation}
where $\tilde p$, $\tilde q$ are operators of the form
\begin{equation}\label{pqtilde}
\tilde p=\left(-\frac{i}{z}\frac{\partial}{\partial\nu}
+\frac{\mu}{2}z\right),\qquad
\tilde q=\left(-\frac{i}{z}\frac{\partial}{\partial\mu}
+\frac{\nu}{2}z\right);
\end{equation}
while $z$, in the space of variables $x,\mu,\nu$ should be 
intended as the derivative with
respect to $x$, i.e.
\begin{equation}\label{zx}
z\leftrightarrow i\frac{\partial}{\partial x}
\end{equation}
and when it appears in the denominator is understood as an 
integral operator.
Furthermore the right arrow over $\Pi$ means that, with respect 
to the order of Eq. (\ref{RP}),
the operators $\tilde p$ and $\tilde q$ act on the right, i.e. on 
$\langle p|:\hat K_{\mu,\nu}:|q\rangle$.
Under the hypothesis of regularity of $w$ on the boundaries, 
we can perform integrations by parts
in Eq. (\ref{Piright}) disregarding the surface terms, to get
\begin{equation}\label{Pileft}
\int dxd\mu d\nu\; \left\{\dot w(x,\mu,\nu,t)+
w(x,\mu,\nu,t){\overleftarrow\Pi}\left({\check p},
{\check q}\right)\right\}
\langle p|:\hat K_{\mu,\nu}:|q\rangle=0\,,
\end{equation}
where now $\overleftarrow\Pi$ means that the operators 
$\check p$, $\check q$ 
\begin{equation}\label{tiltil}
{\check p}=\left(-\frac{i}{z}\frac{\partial}{\partial\nu}
-\frac{\mu}{2}z\right),\qquad
{\check q}=\left(-\frac{i}{z}\frac{\partial}{\partial\mu}
-\frac{\nu}{2}z\right);
\end{equation}
act on
the left, i.e. on the product of coefficients 
$c_{n,m}(-z,\mu,\nu)$ with the marginal distribution
$w$. Finally, using the completness property of the 
Fourier exponents given by
$\langle p|:\hat K_{\mu,\nu}:|q\rangle$
we arrive at the following equation of motion for the 
marginal distribution function
\begin{equation}\label{wmotion}
\partial_t w+w{\overleftarrow\Pi}\left({\check p},
{\check q}\right)=0\,.
\end{equation}
Thus, the distribution function which depends on extra 
parameters obeys 
a classical
equation which preserves the normalization condition of the
distribution. In this sense we always can reduce the quantum 
behaviour of the system to the
classical behaviour of a set of marginal distributions. Of
course, this statement respects the uncertainty relation because 
the measurable marginal
distributions are the distributions for individual variables. 
That is the 
essential difference (despite of
some similarities) of the introduced marginal distribution from 
the discussed quasi-distributions,
including  the real positive Q-function, which depends on the two 
variables of the phase space and
is normalized with respect to these variables.
This our approach to the quantum dynamics remind the 
formulation of quantum mechanics
without the wave functions made in Ref. \cite{Oco}.
In reality, it is not guaranteed (from a rigorous mathematical 
point of view) 
that given the initial condition
for the function $w$ there exist a solution of Eq. 
(\ref{wmotion}), but we demonstrate \cite{ourFofP} the 
existence of an invertible map connecting the Schr\"odinger 
equation with
Eq. (\ref{wmotion}), so there is a one to one correspondence 
between 
their solutions. Then, any solvable quantum systems should admit
an evolutionary solution for $w$. 
For example, a harmonic oscillator in an initial coherent state 
$|\alpha_0\rangle$, gives \cite{ourFofP}
\begin{eqnarray}\label{wtcoh}
&&w(x,\mu,\nu,t)=\frac{1}{\sqrt{\pi}}\left[
\mu^2+\nu^2\right]^{-1/2}
\exp\left[-2|\alpha_0|^2-\frac{x^2}{\nu^2}
+2\sqrt{2}\frac{x}{\nu}\left(
{\rm Re}\{\alpha_0\}\cos\,t
-{\rm Im}\{\alpha_0\}\sin\,t\right)\right]\nonumber\\
&&\times
\exp\left[\frac{1}{\mu^2+\nu^2}\left(
\frac{\mu}{\nu}x
+\sqrt{2}{\rm Re}\{\alpha_0\}(\mu\sin\,t+\nu\cos\,t)
+\sqrt{2}{\rm Im}\{\alpha_0\}(\nu\sin\,t-\mu\cos\,t)
\right)^2\right]\,.
\end{eqnarray}

\section{Comparision between symplectic 
tomography and homodyne tomography}

In this section we connect the discussed realization  
of the tomographic 
principle with the usual one, i.e. the homodyne tomography.
The relations among symplectic tomography and optical homodyne 
tomography are also discussed by Wunsche \cite{Wunsche}.
To get the density operator in terms of the marginal 
distribution, analogously to  Eq.
(\ref{density}), we can start from another operator 
identity such as \cite{cahgla69}
\begin{equation}\label{rho2}
\hat\rho=\int \frac{d^2\alpha}{\pi}\hbox{Tr}\{\hat\rho\hat D 
(\alpha)\}D^{-1}(\alpha)
\end{equation}
which, by the change of variables $\mu=-\sqrt{2}{\rm Im} \,\alpha,
\quad\nu=\sqrt{2}{\rm Re} \,\alpha$,
becomes
\begin{equation}\label{rho3}
\hat\rho=\frac{1}{2\pi}\int d\mu d\nu\quad \hbox{Tr}\{\hat\rho 
e^{-i\hat X}\}e^{i\hat X}
=\frac{1}{2\pi}\int d\mu d\nu\quad \hbox{Tr}\{\hat\rho 
e^{-i\hat x}\}e^{i\hat x}
\end{equation}
where $\hat x=\hat X-\delta$. The trace can be now evaluated
using the complete set of eigenvectors $\{|x\rangle\}$ 
of the operator
$\hat x$, obtaining
\begin{equation}\label{trace}
\hbox{Tr}\{\hat\rho e^{-i\hat x}\}=\int dx\quad 
w(x,\mu,\nu)e^{-ix}
\end{equation}
then, putting this one into Eq. (\ref{rho3}), we have a relation 
of the same form of Eq.
(\ref{density}) with the kernel given by
\begin{equation}\label{kerbis}
\hat K_{\mu,\nu}=\frac{1}{2\pi}e^{-ix}e^{i\hat x}=
\frac{1}{2\pi}e^{-ix}e^{i\mu{\hat q}+i\nu{\hat p}},
\end{equation}
which is the same of Eq. (\ref{kersym}) setting $z=1$. It means 
that we now have 
only one particular Fourier component due to the particular 
change of variables (the most general
should be $z\mu=-\sqrt{2}{\rm Im} \,\alpha$ and $z\nu=
\sqrt{2}{\rm Re} \,\alpha$).

In order to reconstruct the usual tomographic formula for 
the homodyne detection \cite{APL} we
need to pass in polar variables, i.e. $\mu=-r\cos\phi,
\quad\nu=-r\sin\phi$,
then 
\begin{equation}
\hat x\to -r\hat x_{\phi}=-r[\hat q\cos\phi+\hat p\sin\phi].
\end{equation}
Furthermore, denoting with $x_{\phi}$ the eigenvalues of 
the quadrature operator $\hat x_{\phi}$, we
have
\begin{equation}
\hbox{Tr}\{\hat\rho e^{-i\hat x}\}=\hbox{Tr}\{\hat\rho 
e^{ir\hat x_{\phi}}\}=
\int dx_{\phi}\quad w(x_{\phi},\phi)e^{irx_{\phi}}
\end{equation}
and thus, from Eq. (\ref{rho3})
\begin{equation}
\hat\rho=\int d\phi dx_{\phi}\quad w(x_{\phi},\phi)\hat K_{\phi}
\end{equation}
with
\begin{equation}\label{kerAPL}
\hat K_{\phi}=\frac{1}{2\pi}\int dr\quad re^{ir(x_{\phi}-
\hat x_{\phi})}
\end{equation}
which is the same of Ref. \cite{APL}.
Substantially, the kernel of Eq. (\ref{kerAPL}) is given by 
the radial integral of the kernel of Eq.
(\ref{kerbis}), and this is due to the fact that we pass from 
a general transformation, with two free
parameters, to a particular transformation (homodyne rotation) 
with only one free parameter, and then
we need to integrate over the other one.

\section{Photon number tomography}

Here we investigate another possible realization of the 
tomographic 
principle. More precisely
we derive an invariant relation connecting the 
density operator 
of the radiation field
with the number probabilities \cite{oureurophys}. 
So that in this case the measured observable has 
a discrete spectrum.

Indeed Wallentowitz and Vogel \cite{WVogel} 
proposed the 
s-parametrized Wigner function
reconstruction similar, in its essence, to the present 
one by using 
direct photon 
counting, and contemporarely an analogous scheme was adopted by 
Banaszek and W\'odkiewicz \cite{wodkie}.

We named 
the state retrieval by direct photon counting 
"Photon Number Tomography" since it is made by just detecting 
the number 
of photons at a given reference field and then scanning both 
its phase 
and its amplitude; differently from the usual homodyne 
tomography where a 
marginal distribution is recorded by homodyne measurements 
and then 
scanning only the phase.
The method can be used either 
when the 
output beam is mixed with a
reference field at a beamsplitter, as in Refs.
\cite{WVogel,wodkie}, or in
a physical situation similar to the one proposed by Brune {\it et 
al.} \cite{haroche} for the
generation and measurement of a Schr\"odinger cat state, allowing 
thus the 
possibility of
reconstructing its density matrix and, more generally, 
the possibility to 
reconstruct cavity QED field state when there is not a direct 
access to the field \cite{oureurophys}.

As claimed in Ref. \cite{cahgla69}, 
the generalized version of Eq. (\ref{rho1}) is
\begin{equation}\label{denop}
\hat\rho=\int\frac{d^2\alpha}{\pi}W(\alpha,s)\hat T(\alpha,-s)\,,
\end{equation} where the $s$-ordered wheight function 
$W(\alpha,s)$  
may 
be identified with the 
quasiprobability distributions $Q(\alpha)$, $W(\alpha)$ and 
$P(\alpha)$ 
when the ordering parameter
$s$ assumes the values $-1, 0, 1$ respectively; while the 
operator $\hat 
T$ represents the complex
Fourier transform of the $s$-ordered  displacement operator 
$\hat 
D(\xi,s)=\hat D(\xi)e^{s|\xi|^2/2}$,
which can also be written as \cite{cagla1}
\begin{equation}\label{T2}
\hat T(\alpha,s)=\frac{2}{1-s}\hat 
D(\alpha)\left(\frac{s+1}{s-1}\right)^{\hat a^{\dag}\hat a}
\hat D^{-1}(\alpha)\,.
\end{equation} On the other hand the weight function 
$W(\alpha,s)$ is the 
expectation value of the
operator
$\hat T(\alpha,s)$ \cite{cahgla69}, i.e.
$W(\alpha,s)={\rm Tr}\{\hat\rho\hat 
T(\alpha,s)\}$,
then we obtain
\begin{equation}\label{rho}
\hat\rho=\int \frac{d^2\alpha}{\pi}\sum_{n=0}^{\infty}
\frac{2}{1-s}\left(\frac{s+1}{s-1}\right)^n
\langle n|\hat D(\alpha)\hat\rho\hat D^{-1}(\alpha)|n\rangle \hat 
T(-\alpha,-s)\,.
\end{equation} 
Thus, it becomes clear that the weight function $W(-\alpha,s)$ is 
related to the ability 
of measuring the quantity $\langle n|\hat D(\alpha)\hat\rho\hat 
D^{-1}(\alpha)|n\rangle$ by scanning the whole phase space 
\cite{Royer}, 
i.e. by just varying
$\alpha$. Now, we may consider one mode of the radiation, 
whose state 
$\hat\rho$ 
one wants to
reconstruct, contained inside a cavity  and, 
immediately before the 
photon number measurement, a coherent
reference field is "added" \cite{haroche,Glauber63}, 
so that  we may 
recognize 
\begin{equation}\label{P} 
w(n,\alpha)={\rm Tr}\{\hat 
D(\alpha)\hat\rho\hat
D^{-1}(\alpha)|n\rangle\langle n|\}
=\langle n|\hat D(\alpha)\hat\rho\hat 
D^{-1}(\alpha)|n\rangle
\end{equation} 
as the probability to detect $n$ photons after 
the injection of the
reference field $\alpha$.
Again, as an example of the above marginal distribution we consider 
the state of Eq. (\ref{psicat}), for which we have
\begin{eqnarray}\label{wncat}
w(n,\alpha)&=&\frac{1}
{2[1+\cos(2ab)\exp(-b^2)]n!}
\Bigg[ e^{-|\alpha+a+ib|^2}|\alpha+a+ib|^{2n}
+e^{-|\alpha+a-ib|^2}|\alpha+a-ib|^{2n}\nonumber\\
&+&2e^{-|\alpha+a+ib|^2/2-|\alpha+a-ib|^2/2}
{\rm Re}\{(\alpha+a+ib)^n(\alpha^*+a+ib)^n\}\Bigg]\,.
\end{eqnarray}

The addition process we are considering, following Ref. 
\cite{haroche},
"is quite different from 
the combination of fields produced by a beam splitter, which mixes 
together distinct modes coupled to its two ports and introduces 
vacuum 
noise  even in the absence of any classical input field". 
We are indeed
describing a much simpler field amplitude superposition mechanism, 
discussed in the Glauber's pioneering work \cite{Glauber63}.
The photon number distribution (\ref{P}) results as the 
projection of the 
field state $\hat\rho$ over a displaced number state \cite{Knight}.
In this realization the transformation ${\cal G}$ is represented 
by the 
displacement operator, but here the "marginal" distribution is 
a discrete 
probability.

The photon counting could be made either 
by means of atoms \cite{BruneWalls}, as in the case of 
microwave cavity 
field, or by direct detection
of the outgoing optical total field. 
Furthermore, setting 
\begin{equation}\label{K}
\hat K_s(n,\alpha)=\frac{2}{1-s}
\left(\frac{s+1}{s-1}\right)^n\hat T(-\alpha,-s)\,,
\end{equation} we have, from Eq. (\ref{rho})
\begin{equation}\label{rhosample}
\hat\rho=\sum_{n=0}^{\infty}\int\frac{d^2\alpha}{\pi} 
w(n,\alpha)\hat K_s(n,\alpha)\,.
\end{equation} 
Thus, analogously to Eq. (\ref{density}), we may
assert that a density matrix element can be experimentally 
sampled if  
there exist at least one 
value  of the parameter $s$ inside the range $[-1,1]$ for  
which the 
corresponding matrix element
of the kernel operator $\hat K_s$ is bounded. 
From Eqs. (\ref{K}) and (\ref{T2}), one immediately recognizes 
that this 
is possible for $s\in(-1,0]$,
in the number, coherent and squeezed representations and 
not in the 
coordinate basis. For $s=-1$, in Eq. (\ref{K}) $n=0$ only survives,
however as was shown in Ref. \cite{cahgla69}, the 
operator $\hat T$ becomes singular and can be used to construct
an arbitrary density matrix when it is only weighted with a well 
behavied function. It means that $w(0,\alpha)\equiv 
Q(\alpha)$ in Eq. (\ref{rhosample}).

We would stress the fact that the Photon Number 
Tomography, in contrast to optical homodyne tomography, does
not need of sophisticated computer processing of the experimental 
data,
the quantity measured in the experiment is proportional to the 
quasiprobability distribution at a given phase space point. 
In particular our scheme becomes
especially useful in 
the intracavity optical tomography 
where other similar schemes 
\cite{WVogel,wodkie} are not applicable; in fact it can be 
adopted in a situation
in which the photon number is measured indirectly 
using a sequence of atoms
passing through the cavity
\cite{BruneWalls}, with the quantum efficiency, determined 
only by the duration of the measurement process 
(i.e. the length of the 
sequence), that could be very high \cite{haroche}. 
In that case, the 
scheme has also the advantage of being QND. 
Furthermore, it results suitable for cavity QED 
characterization, like Ref. \cite{meysc},
 allowing the reconstruction of 
nonclassical states as well,
 which are extremely sensitive to losses and then 
their detection seems prohibitive  by means of an 
outgoing field as in 
Refs. \cite{WVogel,wodkie}.

The principle of Photon Number Tomography 
turned out to be useful also for the reconstruction
of the motional quantum state of a trapped atom \cite{Leib}.

Finally, a method for direct sampling 
of the density matrix in the Fock basis, 
based on the photon number tomography, 
is implemented in Ref. \cite{welsch},
resulting as an improvment since 
it has been shown that in this case it is sufficient
to vary only the phase of the reference field,
keeping its amplitude constant.

\section{Conclusions}

We have shown that different methods for the state reconstruction 
could be 
ascribed to the same principle. In that sense we have provided 
a sort of 
unification among different (analytical) approaches.
At the same time we have given the physical interpretation of the 
presented realizations and the new observables introduced offer the 
possibility of using various measurements, different from the usual 
ones, to determine the state of the system under study.
The relations between various measurable probabilities are provided 
in a forthcoming paper \cite{marg}.

\section*{Acknowledgments}
V.I.M. deeply acknowledges the Universities of Camerino 
and Naples for kind of hospitality.

\end{document}